\documentclass[11pt,a4paper]{article}%
\usepackage{amsmath}
\usepackage{amsfonts}
\usepackage{amssymb}
\usepackage{graphicx}
\usepackage{color}%
\providecommand{\U}[1]{\protect\rule{.1in}{.1in}}
\pdfoutput=1
\providecommand{\U}[1]{\protect\rule{.1in}{.1in}}
\begin{document}

\title{Causal Structure and Birefringence in Nonlinear Electrodynamics}
\author{C. A. M. de Melo$^{1,2}%
$\thanks{cassius.anderson@gmail.com} , L. G. Medeiros$^{3}$%
\thanks{leogmedeiros@gmail.com} , P. J. Pompeia$^{4,5}$%
\thanks{pedropjp@ifi.cta.br} }
\date{}
\maketitle

\begin{center}
$^{1}$Instituto de Ci\^{e}ncia e Tecnologia, Universidade Federal de
Alfenas.\newline Rod. Jos\'{e} Aur\'{e}lio Vilela (BR 267), Km 533,
n${{}^{\circ}} $11999, CEP 37701-970, Po\c{c}os de Caldas, MG, Brazil.\newline%
$^{2}$Instituto de F\'{\i}sica Te\'{o}rica, Universidade Estadual
Paulista.\newline Rua Bento Teobaldo Ferraz 271 Bloco II, P.O. Box 70532-2,
CEP 01156-970, S\~{a}o Paulo, SP, Brazil.\newline$^{3}$Escola de Ci\^{e}ncia e
Tecnologia, Universidade Federal do Rio Grande do Norte. \newline Campus
Universit\'{a}rio, s/n - Lagoa Nova, CEP 59078-970, Natal, RN, Brazil.\newline$^{4}%
$Inst. de Fomento e Coordena\c{c}\~{a}o Industrial, Departamento
de Ci\^{e}ncia e Tecnologia Aeroespacial. \newline Pra\c{c}a Mal. Eduardo
Gomes 50, CEP 12228-901, S\~{a}o Jos\'{e} dos Campos, SP, Brazil.
\newline$^{5}$ Instituto Tecnol\'{o}gico de Aerona\'{u}tica, Departamento
de Ci\^{e}ncia e Tecnologia Aeroespacial. \newline Pra\c{c}a Mal. Eduardo
Gomes 50, CEP 12228-900, S\~{a}o Jos\'{e} dos Campos, SP, Brazil.\newline
\end{center}

\begin{abstract}

We investigate the causal structure of general nonlinear electrodynamics and determine which Lagrangians generate an effective metric conformal to Minkowski. We also proof that there is only one analytic nonlinear electrodynamics presenting no birefringence.

\end{abstract}


\section{Introduction}

The validity of the Principle of Superposition is one of the cornerstones of
Maxwell Electrodynamics. At macroscopic level and in the vacuum, there are
plenty of experiments which prove such linearity at high confidence level.
Notwithstanding, in material media \cite{Landau} or when high intensity
microscopic fields  \cite{Born, Born-Infeld} or quantum effects
\cite{Euler-Kockel, Euler-Heisenberg, Schwinger}are involved, it is well
known that nonlinear effects can arise. From the theoretical point of
view, extensions of Maxwell theory are proposed to deal with divergences
\cite{Born-Infeld, Podolsky} or they emerge from more general theories, e.g.
higher order gauge theories \cite{Annals} and string theory \cite{Cordas,
Cordas2,Cordas3}.

Nonlinear electrodynamics in material media have been considered as a way to
implement analog black holes \cite{DeLorenci} and many cosmological effects of
nonlinear electrodynamics have been investigated \cite{Breton, Mosquera,
MosqueraNovello}, as well as the impact of these nonlinear fields on the
structure of gravitational black holes \cite{AyonGarcia}.

In this work , we deal with photon propagation on general nonlinear
electrodynamics. There are at least three different approaches to this
problem, due to Boillat \cite{Boillat}, Bialynicka-Birula and
Bialynicki-Birula \cite{Birula} and Novello and collaborators \cite{Novello}.
In all these case, the propagation can be described by means of an effective
metric as a gravitational analog. Here we work with the Boillat-Birula metric
since it is conformally equivalent to Novello's one (see appendix
\ref{NovelloConforme} for more details).

The analysis of photon propagation in nonlinear theories is essential to
determine the causal structure of the theory. Up to now, this has been done
only for a special class of Lagrangians satisfying%
\begin{equation}
\sigma_{1}=\left[  L_{F}^{2}+2L_{F}\left(  GL_{FG}-FL_{GG}\right)
+G^{2}\left(  L_{FG}^{2}-L_{FF}L_{GG}\right)  \right]  \neq0\label{sigma1}%
\end{equation}
with $L=L\left(  F,G\right)  $, $L_{F}\equiv\frac{\partial L}{\partial F}%
$,$\;L_{G}\equiv\frac{\partial L}{\partial G}$, $F\equiv f_{\mu\nu}f^{\mu\nu
},\,G\equiv f_{\mu\nu}f^{\ast\mu\nu},\,f^{\ast\mu\nu}=\frac{1}{2}\epsilon
^{\mu\nu\rho\sigma}f_{\rho\sigma}$ and $f_{\mu\nu}=\partial_{\mu}A_{\nu
}-\partial_{\nu}A_{\mu}$. Here, we extend the study of effective metrics to
include $\sigma_{1}=0$. Using this general formulation for the effective
metric, we deal with the problem of finding which nonlinear electrodynamics
preserve the structure of the lightcone, i.e., what is the form of the
Lagrangians whose effective metric is conformal to the Minkowski one. We find
that there is only one special Lagrangian conformal to Minkowski in all
space-time directions, corresponding to add a topological term to the Maxwell
Lagrangian. Relaxing the analysis to break the isotropy of photon propagation,
we find that there is only a unique analytical Lagrangian which is conformal
to Minkowski in one direction. On the other hand, if analyticity (and
consequently the Maxwell limit) is dismissed, then we proof that it is
possible to construct many Lagrangians conformal to Minkowski space-time.

A second important issue when analyzing nonlinear electrodynamics is the
existence of birefringence. Using our previous results and a series expansion,
we find that there is only one possible analytic Lagrangian for the nonlinear
electrodynamics without birefringence. This Lagrangian is a generalization of
the Born-Infeld, which is recovered demanding that the generalization has a
smooth limit to the Maxwell theory. Here again it is possible to find other no
birefringent electrodynamics only if the Lagrangian is not analytic, implying
in no Maxwell limit. It is important to stress that these results have already
been obtained using a different approach \cite{SG1,SG2}. 

The paper is organized as follows. In section \ref{metric} we review how to
obtain the effective metric and generalize the derivation to include the
condition (\ref{sigma1}). Then we deal in section \ref{conformal} with the
problem of finding which are the possible Lagrangian densities whose effective
metric is conformal to Minkowski and in section \ref{birefringence} we give a
new proof that there is only one analytic Lagrangian function with no
birrefringence. Finally, in section \ref{final} we present our final remarks.

\section{Nonlinear electrodynamics and effective metrics\label{metric}}

The Euler-Lagrange equations for a general nonlinear electrodynamics are%
\[
\partial_{\mu}h^{\mu\nu}=0,~\ \ h^{\mu\nu}\equiv L_{F}f^{\mu\nu}+L_{G}%
f^{\ast\mu\nu}%
\]
with $L_{F}\equiv\frac{\partial L}{\partial F},\;L_{G}\equiv\frac{\partial
L}{\partial G}$.

As in Ref.{\cite{Birula}} we assume that the total electromagnetic field is
composed by strong constant field $F_{\mu\nu}$ and a weak varying wave field
$\phi_{\mu\nu}$, i.e. $f_{\mu\nu}=F_{\mu\nu}+\phi_{\mu\nu}$. In the first
order with respect to $\phi_{\mu\nu}$ the Euler-Lagrange equations become
\[
\left.  \frac{\partial h^{\mu\nu}}{\partial f_{\alpha\beta}}\right\vert
_{f=F}\partial_{\mu}\phi_{\alpha\beta}=0.
\]
Searching for solutions of the form $\phi_{\alpha\beta}=\varepsilon
_{\alpha\beta}e^{ikx}=\left[  k_{\alpha}\varepsilon_{\beta}\left(  k\right)
-k_{\beta}\varepsilon_{\alpha}\left(  k\right)  \right]  e^{ikx}$ the equation
above becomes an algebraic equation. If we decompose $\varepsilon_{\beta
}\left(  k\right)  $ in the basis $\left\{  a^{\mu},\hat{a}^{\mu},k^{\mu
},b^{\mu}\right\}  $, i.e.
\[
\varepsilon^{\mu}=\alpha a^{\mu}+\beta\hat{a}^{\mu}+\gamma k^{\mu}+\delta
b^{\mu},
\]
where $a^{\mu}\equiv F^{\mu\nu}k_{\nu},\,\hat{a}^{\mu}\equiv F^{\ast\mu\nu
}k_{\nu},\,b^{\mu}\equiv F^{\mu\nu}a_{\nu}$, and rewrite the field equations,
we get a pair of equations for $\alpha$ and $\beta$:
\begin{align*}
\alpha\left[  \left(  L_{F}+L_{FG}G\right)  k^{2}+4L_{FF}a^{2}\right]
+\beta\left[  \left(  L_{FF}G-2L_{FG}F\right)  k^{2}+4L_{FG}a^{2}\right]   &
=0,\\
\alpha\left[  4L_{GF}a^{2}+L_{GG}Gk^{2}\right]  +\beta\left[  \left(
L_{F}+L_{GF}G-2L_{GG}F\right)  k^{2}+4L_{GG}a^{2}\right]   &  =0.
\end{align*}
with $a^{2}=\eta_{\mu\nu}a^{\mu}a^{\nu}$. We find that $\gamma$ remains
undetermined and $\delta=0$ (above and henceforth the notation $L_{F}%
\equiv\left(  L_{F}\right)  _{f=F},\,L_{GF}\equiv\left(  L_{GF}\right)
_{f=F},...,$ is considered). Nontrivial solutions are obtained when the
determinant of this system of equations vanish, i.e. when
\begin{equation}
\sigma_{1}\left(  k^{2}\right)  ^{2}+\sigma_{2}k^{2}+\sigma_{3}%
=0,\label{eq2grau}%
\end{equation}
where%
\begin{align}
\sigma_{1} &  =\left[  L_{F}^{2}+2L_{F}\left(  GL_{FG}-FL_{GG}\right)
+G^{2}\left(  L_{FG}^{2}-L_{FF}L_{GG}\right)  \right]  ,\nonumber\\
\sigma_{2} &  =\left[  4L_{F}\left(  L_{FF}+L_{GG}\right)  +8F\left(
L_{FG}^{2}-L_{FF}L_{GG}\right)  \right]  a^{2},\label{sigma 2 dif 0}\\
\sigma_{3} &  =-16\left[  L_{FG}^{2}-L_{FF}L_{GG}\right]  \left(
a^{2}\right)  ^{2}.\nonumber
\end{align}

Eq.(\ref{eq2grau}) is a second order equation for $k^{2}$ and its solution
depends on values assumed by $\sigma_{1},\,\sigma_{2}$ and $\sigma_{3}$. In
Ref.\cite{Birula} the author consider some particular cases, for instance the
cases with $\sigma_{1}\neq0$ where the nonlinear terms are only small
perturbations of the Maxwell theory. Here we extend this analysis and consider
the different solutions obtained for different values of the parameters of the equation.

\subsection{$\sigma_{1}=0,\,\sigma_{2}\neq0$}

First we analyze the case $\sigma_{1}=0,\,\sigma_{2}\neq0$. We obtain a
dispersion relation
\[
k^{2}=\frac{16\left[  L_{FG}^{2}-L_{FF}L_{GG}\right]  }{\left[  4L_{F}\left(
L_{FF}+L_{GG}\right)  +8F\left(  L_{FG}^{2}-L_{FF}L_{GG}\right)  \right]
}a^{2}\Rightarrow g_{eff}^{\mu\nu}k_{\mu}k_{\nu}=0,
\]
with an effective metric given by%
\begin{equation}
g_{eff}^{\mu\nu}\equiv\left[  4L_{F}\left(  L_{FF}+L_{GG}\right)  +8F\left(
L_{FG}^{2}-L_{FF}L_{GG}\right)  \right]  \eta^{\mu\nu}-16\left[  L_{FG}%
^{2}-L_{FF}L_{GG}\right]  F^{\rho\mu}F_{\rho}^{\,\nu}, \label{geff1a}%
\end{equation}
Note that in the limit where the constant field $F_{\mu\nu}$ vanishes, the
effective metric becomes the expected Minkowski metric.

\subsection{$\sigma_{1}\neq0$}

For $\sigma_{1}\neq0$ the solution for Eq.(\ref{eq2grau}) becomes
\[
k_{\pm}^{2}=\frac{2L_{F}\left(  L_{FF}+L_{GG}\right)  -4F\left(  L_{FF}%
L_{GG}-L_{FG}^{2}\right)  \pm2\sqrt{\Delta}}{2L_{F}\left(  FL_{GG}%
-GL_{FG}\right)  +G^{2}\left(  L_{FF}L_{GG}-L_{FG}^{2}\right)  -L_{F}^{2}%
}a^{2}%
\]
where%
\begin{align}
\Delta &  \equiv\sigma_{2}^{2}-4\sigma_{1}\sigma_{3}=\left[  2F\left(
L_{FF}L_{GG}-L_{FG}^{2}\right)  +L_{F}\left(  L_{GG}-L_{FF}\right)  \right]
^{2}+\nonumber\\
&  +\left[  2G\left(  L_{FF}L_{GG}-L_{FG}^{2}\right)  -2L_{F}L_{FG}\right]
^{2}.\label{delta}%
\end{align}
which is also a dispersion relation $g_{eff\pm}^{\mu\nu}k_{\mu}k_{\nu}=0$, but
now two different effective metrics are obtained:%
\begin{align}
g_{eff\pm}^{\mu\nu} &  \equiv\left[  2L_{F}\left(  FL_{GG}-GL_{FG}\right)
+G^{2}\left(  L_{FF}L_{GG}-L_{FG}^{2}\right)  -L_{F}^{2}\right]  \eta^{\mu\nu
}+\nonumber\\
&  -\left[  2L_{F}\left(  L_{FF}+L_{GG}\right)  -4F\left(  L_{FF}L_{GG}%
-L_{FG}^{2}\right)  \pm2\sqrt{\Delta}\right]  F^{\rho\mu}F_{\rho}^{\,\nu
}\label{geff2Conforme}%
\end{align}
In the context of nonlinear electrodynamics the existence of two different
metrics satisfying a dispersion relation indicates that are two geodesics
followed by light (wave perturbation). This is known as the effect of birefringence.

\subsection{$\sigma_{1}=\sigma_{2}=0$}

Finally, for $\sigma_{1}=\sigma_{2}=0$ Eq.(\ref{eq2grau}) is satisfied only if
$\sigma_{3}=0$, i.e.
\begin{align*}
\left[  L_{F}^{2}+2L_{F}\left(  GL_{FG}-FL_{GG}\right)  +G^{2}\left(
L_{FG}^{2}-L_{FF}L_{GG}\right)  \right]   &  =0,\\
\left[  4L_{F}\left(  L_{FF}+L_{GG}\right)  +8F\left(  L_{FG}^{2}-L_{FF}%
L_{GG}\right)  \right]  a^{2}  &  =0,\\
-16\left[  L_{FG}^{2}-L_{FF}L_{GG}\right]  a^{4}  &  =0.
\end{align*}
This system of equations is satisfied only\footnote{This statement is done
supposing that the Lagrangians are real-valued.} if $L=L\left(  G\right)  $
and in this case we should consider not Eq.(\ref{eq2grau}) itself but the
original system that for $\alpha$ and $\beta$. In this case, it is easy to
check that the following equation has to be satisfied:
\[
L_{GG}\left[  Gk^{2}\alpha+\left(  4a^{2}-2Fk^{2}\right)  \beta\right]  =0
\]
For a general ($G-$dependent) Lagrangian and for $\alpha$ independent of
$\beta$ the conditions
\[%
\begin{cases}
k^{2}=0\\
2a^{2}-Fk^{2}=0
\end{cases}
\]
have to be satisfied simultaneously. These are dispersion relations and they
allow us to find the effective metrics:
\[
g_{eff\left(  a,\hat{a}\right)  }^{\mu\nu}=%
\begin{cases}
\eta^{\mu\nu}\\
F\eta^{\mu\nu}-2F^{\rho\mu}F_{\rho}^{\,\nu}%
\end{cases}
\]
This way two effective metrics are found. The metric $g_{eff\left(  a\right)
}^{\mu\nu}=\eta^{\mu\nu}$ is the effective one for the components of
$\varepsilon^{\mu}$ propagating in the $a^{\mu}$ direction while
$g_{eff\left(  \hat{a}\right)  }^{\mu\nu}=F\eta^{\mu\nu}-2F^{\rho\mu}F_{\rho
}^{\,\nu}$ is the effective metric for the components of $\varepsilon^{\mu}$
in the $\hat{a}^{\mu}$ direction.

\section{Conformal metrics\label{conformal}}

Now we turn to the problem of finding those effective metrics that are
conformal to Minkowski. As verified earlier, the effective metric is obtained
from a solution for a second order equation, Eq.(\ref{eq2grau}), which is
parameter dependent ($\sigma_{i},\,i=1,2,3$). To look for the metrics that are
conformal to Minkowski we separate the cases according to possible values
assumed by $\sigma_{1}$.

\subsection{$\sigma_{1}\neq0$}

For $\sigma_{1}\neq0$ we have verified that two effective metrics arise. We
will consider the form given by Eq.(\ref{geff2Conforme}). To obtain a metric
conformal to Minkowski the square brackets multiplying $F^{\rho\mu}F_{\rho
}^{\,\nu}$ must vanish. Two cases are considered below. The first one is the
case where both effective metrics are conformal to Minkowski. The second one
is the case where only one of the effective metrics is conformal.

\subsubsection{Full conformity ($\sigma_{2}=\sigma_{3}=0$)}

The case where both effective metrics are conformal to Minkowski is the one
where both coefficients $2L_{F}\left(  L_{FF}+L_{GG}\right)  -4F\left(
L_{FF}L_{GG}-L_{FG}^{2}\right)  \pm2\sqrt{\Delta}$ are zero. This is only
possible if
\[%
\begin{cases}
\Delta=0\\
2L_{F}\left(  L_{FF}+L_{GG}\right)  -4F\left(  L_{FF}L_{GG}-L_{FG}^{2}\right)
=0
\end{cases}
\]
simultaneously. If we trace calculations back to Eq.(\ref{eq2grau}), it is
straightforward to verify that this corresponds to taking $\sigma_{2}%
=\sigma_{3}=0$.

Once $\Delta$ is composed by the sum of two positive terms - see
Eq.(\ref{delta}) - and we are considering only real-valued Lagrangians then
these two equations split in three, so that
\begin{align}
2L_{F}\left(  L_{FF}+L_{GG}\right)  -4F\left(  L_{FF}L_{GG}-L_{FG}^{2}\right)
&  =0\label{1}\\
2F\left(  L_{FF}L_{GG}-L_{FG}^{2}\right)  +L_{F}\left(  L_{GG}-L_{FF}\right)
&  =0\label{2}\\
2G\left(  L_{FF}L_{GG}-L_{FG}^{2}\right)  -2L_{F}L_{FG}  &  =0 \label{3}%
\end{align}

The first one can be replaced in Eq.(\ref{2}) leading to
\[
L_{F}L_{GG}=0\Rightarrow L_{GG}=0\Rightarrow L=X\left(  F\right)  +GY\left(
F\right)
\]
The condition $L_{F}=0$ is not of our interest since this would lead to
$\sigma_{1}=0$, what is inconsistent with our assumption. Eq.(\ref{3})
becomes
\[
Y_{F}=0\Rightarrow Y=a=constant.
\]

Finally, we backsubstitute this result in Eq.(\ref{1}) and it follows:
\[
2X_{F}\left(  X_{FF}\right)  =0\Rightarrow X=cF.
\]

So, taking $\sigma_{2}=\sigma_{3}=0$, we conclude that the Lagrangians that
lead to metrics that are conformal to Minkowski are of the type
\begin{equation}
L=aG+cF,\label{L1}%
\end{equation}
i.e. Lagrangians that are a linear combination of $F$ and $G$. This Lagrangian
is the sum of the Maxwell Lagrangian and a linear function of $G$. Such
Lagrangian changes the energy-momentum content of the theory but has no effect
on the usual Maxwell field equations due to Bianchi identities.

\subsubsection{Conformity in one direction ($\sigma_{2}\neq0,\,\sigma_{3}=0$)}

The case in which $\sigma_{2}\neq0,\,\sigma_{3}=0$ , i.e. $L_{FG}^{2}%
-L_{FF}L_{GG}=0$, implies $\Delta=\sigma_{2}^{2}$ and we have
\[
\sigma_{2}>0\Rightarrow%
\begin{cases}
g_{eff+}^{\mu\nu} & =\left[  2L_{F}\left(  FL_{GG}-GL_{FG}\right)  -L_{F}%
^{2}\right]  \eta^{\mu\nu}-\left[  4L_{F}\left(  L_{FF}+L_{GG}\right)
\right]  F^{\rho\mu}F_{\rho}^{\,\nu}\\
g_{eff-}^{\mu\nu} & =\left[  2L_{F}\left(  FL_{GG}-GL_{FG}\right)  -L_{F}%
^{2}\right]  \eta^{\mu\nu}%
\end{cases}
\]
For $\sigma_{2}<0$, we have $g_{eff+}^{\mu\nu}\leftrightarrow g_{eff-}^{\mu
\nu}$ so that either if $\sigma_{2}<0$ or $\sigma_{2}>0$ one of the effective
metrics is conformal to Minkowski metric. In this case the conformity is
obtained in only one direction of propagation.

The Lagrangians that satisfy $\sigma_{3}=0$ were analyzed assuming analyticity
at $\left(  F,G\right)  =\left(  0,0\right)  $. A general form of the
Lagrangian is found to be
\begin{equation}
L=f\left(  aF+bG\right)  +cF+eG \label{sol geral1}%
\end{equation}
where $a,\,b,\,c,\,e$ are constants and $D\equiv ae-bc\neq0$ (if $D=0$ then
the particular solution $L=f\left(  aF+bG\right)  $ should be considered).

If we do not claim analyticity at $\left(  F,G\right)  =\left(  0,0\right)  $,
other solutions can be found, for instance,
\[
L=A\left(  bG^{n}+hF^{n}\right)  ^{\frac{1}{n}}+cF+eG\quad\left(
n\in\mathbb{R}^{\ast},\,n\neq1;~\left\{  A,b,h,c,e\right\}  \in\mathbb{R}%
\right)  .
\]

\subsection{$\sigma_{1}=0$ and $\sigma_{2}\neq0$}

For $\sigma_{1}=0$ and $\sigma_{2}\neq0$ as can be seen in Eq.(\ref{geff1a})
conformity to Minkowski is obtained when the coefficient $\left[  L_{FG}%
^{2}-L_{FF}L_{GG}\right]  $ is null. This is equivalent to consider
$\sigma_{3}=0$. As mentioned above, the Lagrangians (analytical at $\left(
F,G\right)  =\left(  0,0\right)  $) that satisfy this condition are those of
the type given in Eq.(\ref{sol geral1}). However, we are also considering
$\sigma_{1}=0$ so an extra condition has to be satisfied:
\begin{equation}
L_{F}^{2}+2L_{F}\left(  GL_{FG}-FL_{GG}\right)  =0,\label{EDP1}%
\end{equation}
with
\begin{equation}
L_{F}\left(  L_{FF}+L_{GG}\right)  \neq0\;\left(  \sigma_{2}\neq0\right)
.\label{sigma 2 igual 0}%
\end{equation}

We define
\[%
\begin{cases}
X\equiv aF+bG\\
Y\equiv cF+eG
\end{cases}
\Rightarrow%
\begin{cases}
F=\frac{1}{D}\left(  eX-bY\right)  \\
G=\frac{1}{D}\left(  -cX+aY\right)
\end{cases}
,\,
\]
and verify which restrictions Eq.(\ref{EDP1}) implies on $L=f\left(  X\right)
+Y$. It is important to note that we are considering $D=ae-bc\neq0$. With this
change of variables Eq.(\ref{EDP1}) becomes:
\[
\left(  af_{X}+c\right)  \left[  af_{X}+c+2\frac{b}{D}\left[  \left(
a^{2}+b^{2}\right)  Y-\left(  ca+be\right)  X\right]  f_{XX}\right]  =0,
\]
Since $f$ \ is only $X-$dependent, if we take the term in the square brackets
to be null then we conclude that this equation has a solution only if
$a^{2}+b^{2}=0$. This implies either $a=\pm ib$ with $b\neq0$ or
$b=a=0\Rightarrow D=0$. If we restrict our analysis to real Lagrangians, then
\[
af_{X}+c=0\Rightarrow L\left(  F,G\right)  =\frac{D}{a}G,~a\neq0.
\]
This means that conformity of the effective metric is obtained only with
Lagrangians that are linear in $G$. However, we have to remember that
Eq.(\ref{sigma 2 igual 0}) also has to be satisfied. This does not occur for
this Lagrangian. So we conclude that there are no effective metrics conformal
to Minkowski for $\sigma_{1}=\sigma_{3}=0$ and $D\neq0$.

Now we will check if $D=0$ leads to a conformal metric. In this case, the
general solution is
\[
L=f\left(  aF+bG\right)  =f\left(  X\right)  .
\]
As before, either we have a complex Lagrangian ($a^{2}+b^{2}=0$) or\
\[
L=f\left(  G\right)  .
\]
This is not compatible with Eq.(\ref{sigma 2 igual 0}). We conclude that it
is not possible to find an analytical Lagrangian (at $\left(  F,G\right)
=\left(  0,0\right)  $) that leads to an effective metric conformal to
Minkowski if $\sigma_{1}=0$ and $\sigma_{2}\neq0$.

\subsection{$\sigma_{1}=\sigma_{2}=0$.}

Finally for $\sigma_{1}=\sigma_{2}=0,$ we have found
\[
\quad g_{eff\left(  a,\hat{a}\right)  }^{\mu\nu}=%
\begin{cases}
\eta^{\mu\nu}\\
F\eta^{\mu\nu}-2F^{\rho\mu}F_{\rho}^{\,\nu}%
\end{cases}
\]
and it is trivial to see that there is conformity to Minkowski in only one direction.

\subsection{Summary of the Lagrangians associated with a conformal metric}

We summarize the results obtained above. There is only one Lagrangian whose
effective metrics are completely conformal to Minkowski: $L=aG+cF$. In this
case, the effective metric is simply
\[
g_{eff\pm}^{\mu\nu}=-c^{2}\eta^{\mu\nu}.
\]

If we consider the case where only one of the effective metrics is conformal
to Minkowski, we find $L=f\left(  aF+bG\right)  +cF+eG$. The effective
conformal metric in this case is
\[
g_{eff}^{\mu\nu}=\left[  2\left(  Fb-Ga\right)  bf^{\prime\prime}-\left(
af^{\prime}+c\right)  \right]  \left(  af^{\prime}+c\right)  \eta^{\mu\nu}.
\]
These are the metrics that preserve the light cone of the Maxwell theory.

\section{Analysis of no-birefringence\label{birefringence}}

Now we turn to the problem of analyzing the Lagrangians that do not present birefringence.

We have already checked that for $\sigma_{1}=0,\,\sigma_{2}\neq0$ there is
only one effective metric, so no birefringence occurs. The equation
$\sigma_{1}=0$ is nonlinear so finding an analytical solution is not trivial.
If we suppose that the solutions are analytical at $\left(  F,G\right)
=\left(  0,0\right)  $, then a series expansion can be proposed. Our analysis
shows that only $L\left(  G\right)  $ Lagrangians satisfies $\sigma_{1}=0$ .
However, these Lagrangians do not obey the constraint $\sigma_{2}\neq0$.
Therefore, it is not possible to find Lagrangians with no
birefringence with $\sigma_{1}=0,\,\sigma_{2}\neq0$.

The other possibility is to consider the case where $\Delta=0$ (when
$\sigma_{1}\neq0$) in Eq.(\ref{geff2Conforme}). Since $\Delta$ is composed by
the sum of two square quantities, we must take
\begin{align}
\left[  G\left(  L_{FG}^{2}-L_{FF}L_{GG}\right)  +L_{F}L_{FG}\right]   &
=0,\label{1a}\\
\left[  2F\left(  L_{FG}^{2}-L_{FF}L_{GG}\right)  +L_{F}\left(  L_{FF}%
-L_{GG}\right)  \right]   &  =0, \label{2a}%
\end{align}
simultaneously. These are nonlinear equations whose solutions are not trivial
to obtain. However it is easy to verify that a proper combination of them
leads to
\[
L_{F}\left[  2FL_{FG}-G\left(  L_{FF}-L_{GG}\right)  \right]  =0.
\]
This equation is satisfied either if $L_{F}=0\Rightarrow L=L\left(
G\right)  $ or if
\begin{equation}
2FL_{FG}-G\left(  L_{FF}-L_{GG}\right)  =0. \label{3a}%
\end{equation}
This is a linear equation and if we suppose $L$ to be analytical at $\left(
F,G\right)  =\left(  0,0\right)  $ then it can be decomposed as a series:
\begin{equation}
L=\sum_{i,k=0}^{\infty}a_{ik}F^{i}G^{k}. \label{serie}%
\end{equation}

Recurrence relations are found to be
\begin{align}
a_{i1}  &  =0,\qquad i=1,2,...\nonumber\\
\left(  k+2\right)  \left(  2i+k+1\right)  a_{i\left(  k+2\right)  }  &
=\left(  i+2\right)  \left(  i+1\right)  a_{\left(  i+2\right)  k},\quad
i,k=0,1,2,... \label{relrecor}%
\end{align}
from which follows\footnote{See Appendix \ref{apB} for more details.}%
\[
a_{m\left(  1+2n\right)  }=0\qquad n=0,1,2,...,\quad m=1,2,...
\]

We notice that there are infinite parameters to be determined if we consider
only the linearized equation. In order to constrain the parameters we have to
substitute this series in one of the nonlinear equations. This procedure is
responsible for establishing new constraints for the parameters. However, the
substitution of a series in a nonlinear equation leads to nontrivial
recurrence relations and not always is possible to find a closed expression.
So we chose to use the series in Eq.(\ref{2a}) and analyze the relations order
by order. In particular, up to fourth-order in the Lagrangian (i.e. $i+k\leq4$
in Eq.(\ref{serie}) ) we find
\begin{align*}
L &  =a_{(0)(0)}+a_{(1)(0)}F+a_{(0)(1)}G+a_{(0)(2)}F^{2}+a_{(0)(2)}G^{2}+\\
&  +\frac{2a_{\left(  0\right)  \left(  2\right)  }^{2}}{a_{\left(  1\right)
\left(  0\right)  }}F^{3}+\frac{2a_{\left(  0\right)  \left(  2\right)  }^{2}%
}{a_{\left(  1\right)  \left(  0\right)  }}FG^{2}+5\frac{a_{\left(  0\right)
\left(  2\right)  }^{3}}{a_{\left(  1\right)  \left(  0\right)  }^{2}}F^{4}+\\
&  +\frac{a_{\left(  0\right)  \left(  2\right)  }^{3}}{a_{\left(  1\right)
\left(  0\right)  }^{2}}G^{4}+6\frac{a_{\left(  0\right)  \left(  2\right)
}^{3}}{a_{\left(  1\right)  \left(  0\right)  }^{2}}F^{2}G^{2}+...
\end{align*}

This is the series of the Lagrangian
\begin{equation}
L=C_{0}+C_{1}\sqrt{1+\frac{F}{2q^{2}}-\frac{G^{2}}{16q^{4}}}+C_{2}%
G,\label{BIgen}%
\end{equation}
when it is expanded up to the considered order and with the proper
identification of constant parameters:%
\begin{align*}
a_{(0)(0)} &  =C_{0}+C_{1},\\
a_{(1)(0)} &  =\frac{C_{1}}{4q^{2}},\\
a_{(0)(1)} &  =C_{2},\\
a_{(0)(2)} &  =\frac{-C_{1}}{32q^{4}},
\end{align*}

The Lagrangian presented in Eq.(\ref{BIgen}) satisfies the condition
$\Delta=0$. In particular if we impose that the Maxwell theory is obtained in
the limit $\frac{F}{q^{2}}\approx\frac{G}{q^{2}}\ll1$, then we are forced to
constrain the parameters $C_{0}$, $C_{1}$ and $C_{2}$ in such a way that the
Lagrangian becomes
\[
L=q^{2}\left[  1-\sqrt{1+\frac{F}{2q^{2}}-\frac{G^{2}}{16q^{4}}}\right]  ,
\]
which is the Born-Infeld Lagrangian. So Eq.(\ref{BIgen}) can be seen as a
generalization of the Born-Infeld Lagrangian once it allows the existence of
non-maxwellian limits. That generalization is the unique possible Lagrangian
to show no birefringence when the condition $\sigma_{1}\neq0$ is taken into
account. Except for the term $C_{2}G$, this result is consistent with the one
obtained by G. Boillat {\cite{Boillat}, S. Deser et al. \cite{SG1}
and J. Plebanski \cite{Plebanski}}. As stated before, from the point of view of
the field equations this term has no contribution, so we are lead to the
same conclusion that Boillat obtained. However, it is important to stress
that this topological term could have observational effects on the
energy-momentum content of the theory.

An important observation of the result presented above is that analyticity at
$\left(  F,G\right)  =\left(  0,0\right)  $ was a key point in the analysis.
If we do not take it into account, other solutions can be found. For instance,
if we admit the solution to be separable in Eq.(\ref{3a})\ we find that the
Lagrangian
\[
L=C_{3}\frac{F}{G}%
\]
satisfies Eqs.(\ref{1a}) and (\ref{2a}) with $C_{3}$ being constant. This is
an acceptable mathematical solution although it does not show a maxwellian limit. This Lagragngian was also found by Plebanski \cite{Plebanski, Birula83}.

\section{Final Remarks\label{final}}

Light propagation on nontrivial background is a known branch of physics
\cite{Plebanski}. In material media it means that the surrounding charges
induce modifications in the equations of motion, which is described by
nonlinearities of the field. In the vacuum, virtual pairs induce analog effects
called the vacuum polarization which occurs above the Schwinger limit
$E_{cr}=B_{cr}=m^{2}c^{2}/e\hbar\approx1.3\times10^{18}V/m\approx
4.4\times10^{13}GeV$. In recent years the light cone conditions derived from
the causal structure of the QED$_{4}$ gauge sector was investigated under the
influence of several kinds of nontrivial backgrounds \cite{Dittrich, Latorre}
using an averaging procedure over polarization states. However, This study excludes
the birefringence phenomena. Here we have studied the causal structure of
general nonlinear effective electrodynamics and found what kind of Lagrangian
functions give rise to an effective metric conformal to Minkowski.

The class of Lagrangians derived here includes the possibility of birefringence
and can be tested experimentally using the data of several experiment as,
\emph{e.g.,} the PVLAS collaboration \cite{PVLAS} in order to constrain their
theoretical parameters. We also proved that there is only one special
Lagrangian (\ref{BIgen}) which present no birefringence and, if the Maxwell
limit is required, it turns to be the Born-Infeld electrodynamics. As a
byproduct we have shown that all the three current approachs to the
electromagnetic effective metric are conformally equivalent to each other.

\subparagraph{Acknowledgements}

CAMM thanks the CBPF and ICRA-Brazil for the hospitality
during a short visit when part of this work was done. LGM acknowledges FAPERN-
Brazil for financial support.

\appendix

\section{Appendix: Conformal Equivalence\label{NovelloConforme}}

The effective metric obtained from
\cite{Boillat} and \cite{Birula} can be compared directly after some
notational translations. However, the effective metric obtained in
\cite{Novello} is only conformal equivalent to the first two. Such difference has its origin on two
different ways to describe the perturbations on the electromagnetic field,
using eikonal/plane wave approximation \cite{Boillat, Birula} or by means of
Hadamard theory \cite{Hadamard, NovelloBook}. This subtlety is important
because the conformal factor connecting Boillat-Birula metric with Novello's
one can vanish in some interesting cases.

Although the physical motivations of each approach are
different, it is possible to show that they are equivalent up to conformal
factors. Using the proper notation, it can be verified that the metrics
presented in Eq.(\ref{geff2Conforme}) and Ref.{\cite{Boillat}} are equivalent
while the one given in Ref.{\cite{Novello},}%
\begin{equation}
g_{\pm}^{\lambda\mu}=\left(  G\left(  L_{GF}+\Omega_{\pm}L_{GG}\right)
+L_{F}\right)  \eta^{\lambda\mu}-4\left(  L_{FF}+\Omega_{\pm}L_{FG}\right)
F^{\mu\nu}F_{\nu}^{\,\lambda}\label{geffMario}%
\end{equation}
is not. In this expression,
\[
\Omega_{\pm}=\frac{-\omega_{2}\pm\sqrt{\omega_{2}^{2}-4\omega_{1}\omega_{3}}%
}{2\omega_{1}},
\]
with%
\begin{align*}
\omega_{1} &  =\left(  L_{GG}^{2}-L_{FG}^{2}\right)  G+2FL_{FG}L_{GG}%
-L_{FG}L_{F}\\
\omega_{2} &  =2L_{FG}^{2}F+\left(  2GL_{GF}+L_{F}+2FL_{FF}\right)
L_{GG}-\left(  L_{F}+2GL_{FG}\right)  L_{FF}\\
\omega_{3} &  =GL_{GF}^{2}+L_{F}L_{GF}-L_{FF}^{2}G+2FL_{FF}L_{GF}%
\end{align*}

Although Eq.(\ref{geff2Conforme}) and Eq.(\ref{geffMario}) are not equivalent,
it is straight to show that they are conformal to each other:
\[
g_{\pm}^{\lambda\mu}=\left(  G\left(  L_{GF}+\Omega_{\pm}L_{GG}\right)
+L_{F}\right)  g_{eff\pm}^{\mu\nu},
\]
so that each of the metrics identified by $+$ and $-$ in Eq.(\ref{geffMario})
is equivalent to one of those given in Eq.(\ref{geff2Conforme}) by a conformal
factor, which are $\left(  G\left(  L_{GF}+\Omega_{+}L_{GG}\right)
+L_{F}\right)  $ and $\left(  G\left(  L_{GF}+\Omega_{-}L_{GG}\right)
+L_{F}\right)  $, respectively. This conformal factor can vanish in some of
the cases studied above, turning the analysis simpler when one uses the metric
$g_{eff\pm}^{\mu\nu}$.

\section{Appendix: Recurrence relations \label{apB}}

The analysis of the recurrence relations Eq.(\ref{relrecor}) shows that the
relations for the odd and even coefficients of the first index of $a_{ij}$ are
separated:\
\[
a_{\left(  2n\right)  \left(  2m\right)  }=\left[  \prod_{j=0}^{n-1}\left(
4n+2m-\left(  3+2j\right)  \right)  \right]  \frac{\left(  2m+2n\right)
!!}{\left(  2m\right)  !!}\frac{1}{\left(  2n\right)  !}a_{0\left(
2m+2n\right)  },\quad n\in%
%TCIMACRO{\U{2115} }%
%BeginExpansion
\mathbb{N}
%EndExpansion
^{\ast};m\in%
%TCIMACRO{\U{2115} }%
%BeginExpansion
\mathbb{N}
%EndExpansion
.
\]%
\[
a_{\left(  2n+1\right)  \left(  2m\right)  }=\left[  \prod_{j=0}^{n-1}\left(
4n+2m-\left(  1+2j\right)  \right)  \right]  \frac{\left(  2m+2n\right)
!!}{\left(  2m\right)  !!}\frac{1}{\left(  2n+1\right)  !}a_{1\left(
2m+2n\right)  },\quad n\in%
%TCIMACRO{\U{2115} }%
%BeginExpansion
\mathbb{N}
%EndExpansion
^{\ast};m\in%
%TCIMACRO{\U{2115} }%
%BeginExpansion
\mathbb{N}
%EndExpansion
.
\]
If the series is supposed to converge then the following convergence
conditions must be imposed:
\[
\lim_{m\rightarrow\infty}\left\vert \frac{a_{0\left(  2m+2n+2\right)  }%
}{a_{0\left(  2m+2n\right)  }}\right\vert <1,\,
\]
for the even part of the series and
\[
\lim_{m\rightarrow\infty}\left\vert \frac{a_{1\left(  2m+2n+2\right)  }%
}{a_{1\left(  2m+2n\right)  }}\right\vert <1,\,
\]
for the odd part of the series.

\end{document}